# Probabilistic Smart Contracts: Secure Randomness on the Blockchain

Krishnendu Chatterjee
IST Austria
Klosterneuburg, Austria
krishnendu.chatterjee@ist.ac.at

Amir Kafshdar Goharshady
IST Austria
Klosterneurbug, Austria
amir.goharshady@ist.ac.at

Arash Pourdamghani
Sharif University of Technology
Tehran, Iran
pourdamghani@gmail.com

*Abstract*—In today's programmable blockchains, smart contracts are limited to being deterministic and non-probabilistic. This lack of randomness is a consequential limitation, given that a wide variety of real-world financial contracts, such as casino games and lotteries, depend entirely on randomness. As a result, several ad-hoc random number generation approaches have been developed to be used in smart contracts. These include ideas such as using an oracle or relying on the block hash. However, these approaches are manipulatable, i.e. their output can be tampered with by parties who might not be neutral, such as the owner of the oracle or the miners.

We propose a novel game-theoretic approach for generating provably unmanipulatable pseudorandom numbers on the blockchain. Our approach allows smart contracts to access a trustworthy source of randomness that does not rely on potentially compromised miners or oracles, hence enabling the creation of a new generation of smart contracts that are not limited to being non-probabilistic and can be drawn from the much more general class of probabilistic programs.

## I. INTRODUCTION

**Smart Contracts**. Smart contracts are programs that formalize and secure relationships between parties that communicate over a public network. Smart contracts were first introduced by Nick Szabo in 1997 [1], much before the advent of cryptocurrencies. However, their popularity is due to their applicability in modern cryptocurrencies, such as Ethereum [2].

**Blockchain**. The blockchain protocol was invented by Satoshi Nakamoto as a means of ensuring consensus about transaction results and ownership of funds in Bitcoin [3]. Nevertheless, this protocol is capable of inducing consensus about the results of any well-defined *deterministic* process. Notably, Bitcoin transactions contain scripts [3, 4] that serve as conditions for using the funds in the transaction.

**Cryptocurrency Smart Contracts**. While Bitcoin scripts are useful, they are limited to a few basic operations. Several newer cryptocurrencies allow programs of arbitrary (Turing-complete) complexity, which can be used to implement more complex financial contracts [5-7]. These programs can receive, hold and transfer money in form of cryptocurrency units. Therefore, they are also called *smart contracts*. Currently, the most widely-used platform for smart contracts is Ethereum [2], which is also the second largest cryptocurrency by market cap [8]. In the sequel, we use the terms "program", "smart contract" and "contract" interchangeably. We also use the term "random" to mean pseudorandom and assume that every node of the network can generate local pseudorandom numbers.

**Determinism**. For smart contracts to be enforceable, the entire cryptocurrency network should reach a consensus about their state of execution and the resulting monetary transactions. Basically, if every node of the network has the same understanding about program semantics, then when a function call (change) to the program is added to the blockchain, the node runs the program and computes its resulting state and transactions, which must be the same as those computed by any other node of the network. Ethereum and other programmable cryptocurrencies ensure this property by disallowing non-determinism and randomness in their smart contracts [2].

**Randomness**. While on today's platforms, smart contracts must be non-probabilistic, it is well-known that probabilistic programs are a much richer class than simple non-probabilistic programs [9]. Moreover, there are many real-world financial contracts in which randomness plays a vital role, e.g. casino games and proof-of-stake protocols. See Section III for a discussion of motivating examples. Hence, the lack of probability is a significant limitation for smart contracts.

**Previous Approaches**. Given the necessity of randomization in many real-world applications, several workarounds have been developed for generating pseudorandom numbers to be used in smart contracts. These include using the hash of the current block as a seed, relying on an external provider through an oracle, or creating a smart contract in which anyone can submit randomly-generated numbers and other smart contracts can rely on as a library. See Section IV for more details. However, these approaches suffer from security and incentive issues and cannot be trusted for generating random numbers with possibly enormous financial consequences.

**Security Issues**. Most of the previous approaches give unfair advantages to either the miners or the owners of an oracle. In the worst case, this can lead to a complete centralization of the process of "random" number generation. In other cases, these vulnerabilities allow specific parties to manipulate the process. Hence, a smart contract using these approaches as its source of randomness is automatically trusting either the miners, or the owners of an oracle/other smart contract; both of whom are often anonymous entities.

**Incentive Issues**. Several of the previous approaches rely on random input from multiple participants in order to generate random numbers for use on the blockchain. While these approaches provide participation incentives, they fail to create an incentive for the participants to submit *random* numbers. A participant can get the same rewards by submitting a constant number, e.g. 0, every time. Hence, they rely on participants' honesty, i.e. the assumption that a majority or at least some of the participants are truly submitting random inputs. This is much less secure than relying on incentives, i.e. ensuring that it is in everyone's best interest to submit random inputs, especially given that generating random inputs is a (slightly) more expensive task than reusing a constant.

***Our Contribution and Novelty***. In this paper, our contributions are as follows:

- We study several previous approaches and show that all of them suffer from either security vulnerabilities, or incentive issues, or both.
- In order to enable a rigorous theoretical treatment, we provide a formalization of the requirements for secure random number generation on the blockchain.
- We provide the first secure and well-incentivized approach for generating random numbers on the blockchain. We do this by defining a game on the blockchain that incentivizes its players to play randomly. *This is the first approach that relies on incentives, and does not make any assumption about participants' honesty*. We *prove* that our approach satisfies the formal requirements for security, functionality and incentives.
- Finally, we provide an implementation of our approach as an Ethereum smart contract, showing its applicability to real-world blockchains.

## II. PRELIMINARIES

We now review and define some necessary concepts in blockchain theory, probabilistic programs, and game theory. Our treatment of the blockchain protocol is high-level. See [2, 3, 10] for a more formal discussion.

***Transactions***. The atomic unit of communication in a cryptocurrency network is a *transaction* [3]. A transaction can transfer money or invoke other well-defined behavior [11].

***Blockchain [3]***. A *blockchain* is a distributed ledger of blocks, each containing several transactions. It starts from a predefined initial block and each block in the chain must have a pointer to its predecessor. Hence, the blockchain is essentially a singly-linked list of blocks. Each node of the network keeps a local copy of the blockchain. A transaction is considered to be executed/approved iff it appears in some block of the blockchain. To ensure a consensus about the contents of the blockchain, a node cannot simply add a new block. Instead, blocks must be *mined*.

***Mining and Proof-of-Work***. *Mining* is the process by which a unique node of the network is chosen and given the permission to add a new block to the blockchain [3]. The most common protocol for mining is *proof-of-work* [12], in which a hard computational puzzle is set and the first person to solve it gets the right to add a new block. Network nodes who take part in solving the puzzle are called *miners*. If several miners find solutions at approximately the same time, then a fork happens in the blockchain, i.e. a situation in which there are more than one valid chains. In these cases, the protocol dictates that the longest chain is the consensus chain. Due to the probabilistic nature of mining, one chain will eventually become longer and hence consensus will be reached.

***Mining Incentives***. Proof-of-work mining needs considerable computational resources and electricity. Hence, miners should be incentivized to perform it. There are two incentives for mining [3]: (i) each miner receives a fixed *block reward* in form of cryptocurrency units, for each block she adds, and (ii) transactions can include *transaction fees* that are paid to the miner who adds them to the blockchain. Incentive (i) also serves as the mechanism for creating new currency units.

***Programmable Blockchains [2, 13]***. In a programmable blockchain, transactions can do more than simple money transfers. Specifically, there is a special type of transaction that creates a smart contract, which is a program consisting of several functions and its own dedicated memory (storage). When a smart contract is created, i.e. when its code is stored on the blockchain, it can receive, manage and transfer money in form of cryptocurrency units. However, the smart contract code remains unchangeable and one can only interact with it by calling its functions. Calling functions is also considered a special type of transaction. Smart contracts can also interact with each other, i.e. a smart contract that is called by a network node can itself create or call functions from another contract. Intuitively, when a function call transaction is added to the blockchain, all the nodes of the network execute the relevant smart contract using the parameters provided in the transaction and reach a consensus about its results.

***Ethereum [2]***. Ethereum was the first cryptocurrency to allow Turing-complete smart contracts. It is also the most widely-used platform for smart contracts.

***Ethereum Virtual Machine [10]***. For Ethereum contracts to be well-defined, they must be written in a specific bytecode format that can be executed on the Ethereum Virtual Machine, which is a stack machine designed to ensure that all nodes of the network agree on the execution results. However, there are many languages for writing Ethereum contracts, most notably Solidity [14]. Contracts written in these languages are first compiled into bytecode and then published on the blockchain.

***Determinism and Lack of Randomness [2, 10, 15, 16]***. To ensure that all nodes can reach a consensus about the results of function calls to contracts, Ethereum bytecode does not have any support for non-determinism or randomness.

***Probabilistic Programs [9]***. Extending imperative programs with randomization, i.e. allowing them to access a source for random number generation, leads to the much richer class of probabilistic programs. Probabilistic programs are very widely studied (e.g. see [9, 17-21]) and have many applications in different fields, such as machine learning [22-25], randomized algorithms [26], and analyzing stochastic networks [27-29].

***Probability vs Non-determinism***. In programming languages theory, probability and non-determinism are two different and often orthogonal concepts [30]. The prerequisite for consensus in contracts is that they cannot be non-deterministic. As we show in this work, this does not necessarily mean that they should also be non-probabilistic. We propose a game-theoretic approach that can provide randomness to smart contracts and allow them to be probabilistic programs.

*Value of Contracts*. At the time of writing, there are over a million contracts on Ethereum alone, holding billions of dollars of funds [31]. Hence, formal verification of smart contracts and blockchain protocols is very important and well-studied [32-37]. Similarly, it is vital to ensure that the generation of random numbers is secure, given that any attack on this process might have huge financial consequences.

*Probability Distributions*. Given a finite set $X = \{x_1,\ldots,x_k\}$, a probability distribution on $X$ is a function $\delta : X \to [0,1]$ such that $\delta(x_1)+\ldots+\delta(x_k)=1$. We denote the set of all probability distributions on $X$ by $\Delta(X)$.

*One-shot Games [38]*. A *one-shot game* with $n$ players is a tuple $G = (S_1, S_2,\ldots,S_n, u_1, u_2,\ldots,u_n)$ where:
- Each $S_i$ is a finite set of *strategies* for player $i$ and $S = S_1 \times S_2 \times \ldots \times S_n$ is the set of all *outcomes*; and
- Each $u_i$ is a *utility function* of the form $u_i : S \to \mathbb{R}$.

In a play of the game, each player $i$ chooses one strategy $s_i \in S_i$. The choices are simultaneous and independent. Then, each player $i$ is paid a *utility* of $u_i(s_1, s_2,\ldots,s_n)$ units.

*Mixed Strategies [38]*. A *mixed strategy* $\sigma_i$ for player $i$ is a probability distribution over $S_i$. Intuitively, a mixed strategy $\sigma_i$ is a recipe for player $i$ in order to randomly choose one of the strategies in $S_i$ and play it. A mixed strategy profile is a tuple $\sigma = (\sigma_1, \sigma_2,\ldots,\sigma_n)$ consisting of one mixed strategy for each player. The expected utility $u_i(\sigma)$ of player $i$ in a mixed strategy profile $\sigma$ is defined as $u_i(\sigma) = \mathbb{E}[u_i(s_1, s_2,\ldots,s_n)]$ where each $s_i$ is sampled according to $\sigma_i$.

*Nash Equilibria [39]*. A *Nash equilibrium* of a game $G$ is a mixed strategy profile $\sigma$, such that no player has an incentive to change her mixed strategy $\sigma_i$, assuming that she knows the mixed strategies of the other players. We define $\sigma_{-i}$ as a tuple consisting of all the mixed strategies in $\sigma$ except $\sigma_i$. Formally, $\sigma$ is a Nash equilibrium iff for all $\tilde{\sigma}_i \in \Delta(S_i)$ we have $u_i(\sigma) \geq u_i(\tilde{\sigma}_i, \sigma_{-i})$. A seminal result by John Nash is that every finite game $G$ has a Nash equilibrium [39].

In game theory, Nash equilibria are the standard notion of stability and self-enforceability for non-cooperative games [38], i.e. games in which each player only aims to maximize his own utility. When considering games on the blockchain, the players are pseudonymous and their real identities are unknown. Hence, some players might indeed be the same person and hence cooperate. In these cases, the notions of strong and quasi-strong Nash equilibria are useful.

*Strong Nash Equilibria [40, 41]*. A *strong Nash equilibrium* is a mixed strategy profile in which no group of players have a way to cooperate and change their mixed strategies in order to increase the utility of every member of the group. Formally, $\sigma$ is a strong Nash equilibrium if for any non-empty set $P$ of players and any strategy profile $\tilde{\sigma}_P$ over $P$, there exist a player $p \in P$ such that $u_p(\sigma) \geq u_p(\tilde{\sigma}_P, \sigma_{-P})$.

```
uint endTime, p;
mapping (uint => address) participants;
uint n = 0;
function buyTicket() payable public
{
  if(msg.value==p && block.timestamp<=endTime)
  {
    n = n+1;
    participants[n]=msg.sender;
  }
}

function draw() public
{
  if(block.timestamp>endTime)
  {
    uint x = random(1,n);
    address winner = participants[x];
    winner.send(this.balance);
  }
}
```

**Figure 1. A Lottery Contract**

When considering strong Nash equilibria, the assumption is that the players cannot share or transfer their utilities, so a player agrees to the change of strategies only if his own utility is strictly increased. However, if the players can share utilities, then a group can defect as long as their *total* utility increases, and hence a stronger notion of equilibrium is necessary.

*Quasi-strong Nash Equilibria*. A *quasi-strong Nash equilibrium* is a mixed strategy profile $\sigma$ such that for any non-empty set $P$ of players and any strategy profile $\tilde{\sigma}_P$, it holds that $u_P(\sigma) \geq u_P(\tilde{\sigma}_P, \sigma_{-P})$, where $u_P$ is the sum of utilities of all members of $P$.

In games that are played on the blockchain, quasi-strong equilibria are the right notion for ensuring stability and self-enforceability, because several players might be controlled by the same person and can naturally share their revenues and defect as long as their total revenue increases.

*Remark*. By definition, every quasi-strong equilibrium is also a strong equilibrium, and every strong equilibrium is also a Nash equilibrium. Hence, quasi-strong equilibria are the strongest notion of stability, i.e. if a game has a quasi-strong equilibrium, there are strong guarantees that no group of rational players would defect from the equilibrium.

III. MOTIVATING EXAMPLES

In this section, we provide several motivating examples that highlight the need for randomization in smart contracts.

*Lotteries*. The global lottery industry sells more than 250 billion dollars' worth of tickets each year [42]. Given the huge profit margins in most lotteries, the organizers' revenues are high and the returns for participants are much lower than they could potentially be if the lotteries were implemented using smart contracts instead of trusted third-parties [43].

A lottery is a simple probabilistic contract. It consists of two phases. In the first phase, anyone can buy a ticket by paying a predefined amount $p$. We assume that $n$ people buy tickets. In the second phase, a random number $x$, between *1* and *n*, is drawn and the entire money is paid to person *x*. Hence, a lottery can be implemented as shown in Figure 1. Note that the randomization line shown in bold is crucial, but not natively available to smart contract programmers.

*Gambling and Casino Games*. Lotteries are the simplest form of gambling. The global gambling market is expected to reach revenues of over 525 billion dollars by 2023 [44]. In modern casinos, there are many more complex and interactive gambling games, e.g. roulette [45]. These games can also be implemented by probabilistic programs [21]. Given that Ethereum is Turing-complete, one can implement any set of game rules in its smart contracts. However, random number generation is still a necessary step because these games are inherently probabilistic.

*Proof-of-stake Mining*. Proof-of-work uses considerable electricity and has a huge carbon footprint [46]. It was reported in 2014 that Bitcoin mining alone uses more electricity than all of Ireland [47]. Hence, several alternative mining protocols have been proposed [48-52]. A notable approach is proof-of-stake and Ethereum is expected to switch to it in near future [53]. In proof-of-stake protocols, a member of the network is randomly chosen to add the next block to the blockchain and one's chance of being chosen is proportional to the number of coins she holds (her stake).

A method for producing reliable random numbers on the blockchain can be applied in proof-of-stake protocols for randomly choosing the miner who gets to add the next block.

*Proof-of-stake DAO's*. Decentralized autonomous organizations (DAO's) are organizations whose rules are defined by smart contracts and whose shareholders can manage their funds and the decisions of the organization by voting in the contract [54]. If reliable random numbers are generated on the blockchain, then such organizations can use proof-of-stake schemes for randomly choosing a subset of members to vote every time the organization needs to make a decision. In contrast, in current DAO's the voting procedure takes a long time because every action should be put to a vote among all shareholders [54]. DAO's also suffer from a lack of participation by the shareholders in the voting procedure, given that it takes time and energy to vote [55].

## IV. Previous Approaches

In this section, we review several previous approaches to random number generation and show that they suffer from security vulnerabilities and incentive misalignments.

*Using Block Hash or Timestamp*. The simplest approach is to use one of the attributes of the block containing a transaction, e.g. its hash or timestamp, as a seed for generating random numbers in that transaction's execution. In this approach, it is assumed that no party can control the hash of a block or the exact time in which it is mined, hence the random number generation is tamper-proof. This is known to be a very vulnerable, but unfortunately widely-used, approach [56, 57].

One of its vulnerabilities is that it gives undue advantage to the miners. Consider the lottery of Figure 1, if the miner who mines the block containing the call to the `draw` function is a lottery participant and the block timestamp is used as a seed, he can manipulate the timestamp to always win the lottery. If the seed is not manipulatable, e.g. the block hash, then if the miner realizes that he is going to lose the lottery, he can simply ignore the block and decide not to publish it on the network. By doing so, he loses his block reward, but gains an extra chance of winning the lottery. If a huge amount of money is at stake, then this strategy is rational. Note that most lotteries have millions of participants, so no real-world lottery can be securely implemented using this approach. Concretely, if the amount of money at stake is more than the block reward, then the miners should not be trusted in generating seeds for random numbers. A well-known example of such methods is the Ethereum Lottery project [58], which was suspended due to potential tampering by the miners.

*Oracles [59]*. Another approach is to use an Oracle [56, 60]. By design, smart contracts can only access data that is written on the blockchain [10]. Oracles are third-party services that access outside sources and write the obtained data on the blockchain so that smart contracts can use it. The data usually comes with a signature from the oracle, promising that it was collected from a predefined source. To generate random numbers, one can create an oracle, e.g. using the oraclize service [60], that obtains random numbers from an outside source, e.g. random.org [61], and puts them on the blockchain. However, this approach requires trusting the oracle owners and gives them the power to report any number as the "random" output. Hence, it centralizes the random number generation process and cannot be considered secure.

*Using Commitment Schemes [56, 62, 63]*. In this classical approach, all participants of a contract create a random number together. The approach works in two steps. In the first step, each participant generates her own random number $x$ and then commits to $x$ by sending the hash of $x$, together with a deposit, to the contract. The first step ends after a predefined amount of time. In the second step, each participant must announce the actual random number she has generated by sending it to the contract. The contract checks whether the submitted random number has the right hash. If it does not, the contract confiscates the deposit and ignores this participant's entry. Assuming hash functions are one-way, no participant can change his number after having committed to it. Finally, the second step ends after a predetermined time and then the contract computes the XOR of all valid entries and uses it as the output random number. Note that even if one of the participants honestly generates and provides random numbers, the final output would be random.

We now analyze the security of this approach: A malicious participant can wait for others to uncover their random numbers in the second step. Then, based on other people's numbers, he has the choice of either uncovering his random number or not. This gives him an advantage, e.g. in the lottery example of Figure 1, he gets an extra chance at winning. However, it costs him his deposit and if the deposit is large enough, the approach is secure against this attack. On the other hand, a miner can manipulate the process by not processing some participants' transactions in the second step. To avoid this attack, the second step should be made long enough to ensure that all participants have a chance to add their uncovered numbers to the blockchain.

While this approach is secure, it is not desirable for two reasons. First, it requires the participants to consistently interact with the contract in order to generate their own random numbers. Hence, it cannot be used when the participants are not willing or able to follow this protocol, e.g. it is infeasible for most lottery participants [58]. Second, it has incentive issues, i.e. it provides no incentives to the participants to submit *random* numbers. A participant can get the same results by always submitting a constant, e.g. 0. On the other hand, the protocol relies on the assumption that at least one participant honestly submits random numbers.

*RANDAO and Quanta*. An approach for solving the interaction problem of using commitment schemes is provided by a smart contract called RANDAO [64] and is also used on the Ethereum-based lottery platform Quanta [43]. The RANDAO smart contract acts as a library that provides other contracts with random numbers. These are generated by volunteer participants who interact with the RANDAO contract in a manner similar to the commitment schemes approach. Other contracts pay RANDAO for the generated numbers and RANDAO uses these payments as participation incentives, i.e. pays the participants who submit and uncover inputs correctly. As in the previous approach, each participant has to provide a deposit which will be confiscated if his number is not uncovered in the second step.

RANDAO does not fix the incentive problem and, quite ironically, introduces new security vulnerabilities. First, the numbers generated in RANDAO are often reused in several client contracts. This is a well-known malpractice and vulnerability [65-68]. Second, the confiscated deposits are paid to the owner of RANDAO, hence giving him undue advantage. The owner can participate in the random number generation and decide not to uncover his submission, so as to increase his winning odds in client contracts, all without any consequences. Hence, RANDAO has the same problem as an oracle. It is secure only if one trusts its anonymous developers.

Based on the discussion above, previous approaches for random number generation suffer from various security and incentive issues. In this work, we present a novel game-theoretic approach that rectifies these issues. Note that we are considering real-world programmable blockchains with deterministic non-probabilistic semantics, such as Ethereum, as the environment in which random numbers must be generated. Proposing alternative mining protocols that support random number generation is another active area of research. See [69] for a survey of some attempts in this direction.

## V. OUTLINE OF OUR APPROACH

In the sequel, we provide a novel game-theoretic approach for securely generating random numbers on the blockchain. To do so, we take the following steps:

- We first formally define the requirements that should be satisfied in order for an approach to be considered functional, safe and unmanipulatable (Section VI).
- Then, we define a game in which the only quasi-strong equilibrium is when every player plays uniformly at random (Section VII). Intuitively, one can deploy this game on the blockchain and use the strategies played by its players to generate randomness, because the only stable rational strategy for each player is to play uniformly at random. *This is a key novelty of our approach and ensures that, unlike previous methods, we provide economic incentives guaranteeing that honesty is in every player's best interest, instead of assuming that the players remain honest due to their goodwill.*
- We use the game above to define a random bit generation contract, which is the heart of our approach (Section VIII) and can be used as a library that provides secure randomness to other contracts.
- We meticulously analyze our approach (Section IX) and prove that it satisfies all the formalized requirements.
- Finally, we provide a proof-of-concept implementation of our approach for Ethereum (Section X), showing its applicability in real-world programmable blockchains.

## VI. SPECIFICATION OF REQUIREMENTS

In this section, we formalize the requirements that must be satisfied by a blockchain random number generation protocol in order to ensure that it is usable, unmanipulatable and secure, and provides the right incentives to all participants.

*Functional Requirements*. The protocol must support the following functionalities:

a) In order to be applicable in real-world blockchains, it must be implementable as a smart contract.
b) Other smart contracts/nodes should be able to use it as a library. Concretely, others should be able to request random bits from the protocol by paying a fee and specifying a deadline, as well as an upper bound $v$ on the potential economic consequences that might arise if the generated random bit is tampered with[1].
c) It may rely on its own participants for generating random bits, but it must not rely on the client.

*Security Requirements.* The protocol must provide the following guarantees:

a) When a request for a random bit is submitted, the client must receive one of the following responses before the deadline:
   i) *Success*: An unmanipulated and untampered random bit, which must be generated after the request and not be known or predictable at the time of the request; or
   ii) *Penalty*: A possibly manipulated random bit, also generated after the request, together with a penalty of at least $v$ units (This ensures that if the random bit is manipulated, its economic consequences can be rectified by the $v$ units of penalty that is paid to the client); or
   iii) *Failure*: A notification of failure to generate a random bit[2] and a full refund.
b) *Randomness guarantee*: The final output bit must be uniformly random as long as at least one of the participants submits a uniformly random input.
c) *Openness:* To avoid centralization, anyone should be able to join as a participant in the protocol.
d) *Safety against malicious miners*: No miner should be

---
[1] Naturally, a more secure random bit is expected to be costlier.
[2] This can happen if the paid fee is too small or the set deadline is too early and hence no one participates in the random number generation protocol.

able to affect the output of the protocol, either by tampering with or withholding the blocks that were used in the random bit generation process.
e) *Safety against malicious participants*: If the protocol is successful, it should be guaranteed that no participant in the random number generation process could have tampered with the output.
f) *Avoiding reuse*: No generated random bit should ever be reused and a new dedicated random bit should be generated for each request.

***Incentive Requirements***. Finally, the protocol must guarantee that there are incentives for the participants to act honestly:
a) No participant should be able to manipulate the output without paying at least $v$ units of penalty, or without changing the output type from *success* to *penalty*.
b) Participants should be incentivized to submit uniformly randomly generated bits as inputs to the protocol, i.e. providing random inputs should be a quasi-strong equilibrium for the participants.

## VII. RANDOM BIT GENERATION GAME (RBG)

We now define a special game which will be used to ensure that our protocol has the right incentive structure.

**RBG.** A *Random Bit Generation game* (RBG) with $n$ players is a game $G$ in which:
- For a player $i$, $S_i = \{0,2\}$ if $i$ is even and $S_i = \{1,3\}$ if $i$ is odd;
- At an outcome $s = (s_1, s_2, \ldots, s_n)$ of the game, the utility of player $i$ is defined as $u_i(s) := \sum_{j \neq i} f(s_i, s_j)$ where

$$f(s_i, s_j) := \begin{cases} 1 & s_i \equiv s_j + 1 \pmod 4 \\ -1 & s_i \equiv s_j - 1 \pmod 4 \\ 0 & \text{otherwise} \end{cases}.$$

Intuitively, in an RBG game, all players choose their actions simultaneously. Then, for every pair $(i, j)$ of players with different parity, a minigame is simulated using their chosen actions, in which a player wins if and only if her chosen number is one more than that of her opponent (Figure 2). A win leads to 1 point and a loss to −1. A player's utility in the game is defined as the sum of her points in the minigames.

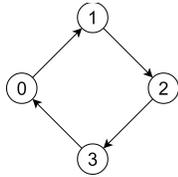

**Figure 2. An RBG Minigame.** An arrow x ⟶ y means a player playing y wins against a player playing x.

***Theorem 1 (Quasi-strong Equilibrium of an RBG.)*** Let $G$ be an RBG game with at least two players and let $\bar{\sigma}$ be a mixed strategy profile defined by $\bar{\sigma}_i = (0.5, 0.5)$ for all $i$, i.e. the mixed strategy profile in which each player $i$ chooses a strategy in $S_i$ uniformly at random. Then, $\bar{\sigma}$ is *the only* quasi-strong equilibrium of $G$.

*Proof.* Let $P$ be a subset of players in $G$. Then, the overall utility of members of $P$ in a strategy profile $\sigma$ is $u_P(\sigma) = \mathbb{E}^{\sigma}\left[\sum_{i \in P} \sum_{i \neq j \pmod 2} f(i, j)\right]$. Note that if both sides of a minigame are in $P$, that minigame only transfers points inside $P$ and has no effect on the overall utility of $P$. So,

$$u_P(\sigma) = \mathbb{E}^{\sigma}\left[\sum_{i \in P} \sum_{j \notin P \wedge i \neq j \pmod 2} f(i, j)\right]$$
$$= \sum_{i \in P} \sum_{j \notin P \wedge i \neq j \pmod 2} \mathbb{E}^{\sigma}\left[f(i, j)\right]. \quad (\dagger)$$

For all $i$ and $j$, we have $\mathbb{E}^{\bar{\sigma}}[f(i,j)] = 0$. Therefore, $u_P(\bar{\sigma}) = 0$ for all $P$. Suppose that the players in $P$ decide to defect and use a new strategy profile $\sigma_P$ instead of $\bar{\sigma}_P$. Since every player $j \notin P$ still follows $\bar{\sigma}$ and plays uniformly at random, $\mathbb{E}^{\sigma_P, \bar{\sigma}_{-P}}[f(i,j)] = 0$, so $u_P(\sigma_P, \bar{\sigma}_{-P}) = 0 \leq u_P(\bar{\sigma})$. Therefore, $\bar{\sigma}$ is a quasi-strong equilibrium.

We show that no other quasi-strong equilibrium exists. Note that the game $G$ is zero-sum, i.e. utilities of the players always sum to zero. Let $\sigma$ be a quasi-strong equilibrium, and $e$ an arbitrary even-numbered player. Also, let $R$ be the set of all remaining players. If $\sigma_e = (p_0^e, p_2^e) \neq (0.5, 0.5)$, then $e$ plays one of her actions more often than the other. W.l.o.g. let us assume that $p_0^e > p_2^e$. Then for every odd-numbered player $o \in R$, $\sigma_o = (p_1^o, p_3^o) = (1, 0)$, because otherwise the players in $R$ can defect by setting $\sigma_o = (1,0)$ and increase their total utility, which is in contradiction with $\sigma$ being a quasi-strong equilibrium. So every $\sigma_o$ is $(1,0)$. However, this is a contradiction, too, because now $e$ can profit by defecting and changing her strategy to $(p_0^e, p_2^e) = (0,1)$. This contradiction shows that in every quasi-strong equilibrium $\sigma$ we must have $\sigma_e = (0.5, 0.5)$, for every even-numbered player $e$. A similar argument can be applied to odd-numbered players. Therefore, the only quasi-strong equilibrium of the game $G$ is $\bar{\sigma}$. ∎

Intuitively, this theorem means that if an RBG game is played with pseudonymous players on the blockchain, with the possibility that some players be controlled by the same person, then the only stable rational strategy for each player is to play uniformly at random. Hence, an RBG game on the blockchain incentivizes rational players to generate uniform random bits. Even-numbered players generate random bits by playing either 0 or 2 and odd-numbered players generate them by playing either 1 or 3.

## VIII. RANDOM BIT GENERATOR CONTRACT (RBGC)

We now present a novel approach based on RBG games for generating tamper-proof pseudorandom numbers on the blockchain. Our approach combines the strengths of previous approaches, while avoiding their security and incentive vulnerabilities. It consists of a single-instance smart contract, which is called the Random Bit Generator Contract (RBGC).

The RBGC accepts requests for random bit generation from other contracts or network nodes. After receiving a request, it starts the process of generating a dedicated random bit for that request. Anyone can participate in this process by first submitting the hash of their chosen random bit, hence committing to it, and then revealing it. The participants are rewarded for submitting inputs, but the amount of their rewards is determined through an RBG game which incentivizes them to submit uniformly random bits. Moreover, any participant who fails to reveal his choice will be penalized by confiscation of his deposit. We now formalize these steps[3].

**Step 1: Request.** The process of random bit generation always begins with a request from another contract or node. A request is lodged with the RBGC by calling its `requestRandomBit` function, which receives the following parameters:
- A fee $\varphi$, serving as a payment for generating the random bit, whose value is chosen and paid by the client.
- A timestamp $t$, also set by the client, that serves as a deadline for the random bit generation process.
- A value estimation $v$, which is an upper bound on the potential economic consequences that might arise for the client if the provided bit is manipulated.

Upon receiving these, the RBGC rejects the request if the deadline $t$ is too close, i.e. if the total time available for generation of the random bit is less than a predefined constant $t_{min}$. Otherwise, it assigns a request identification number `id` to the request and returns the `id` to the caller of the `requestRandomBit` function. It also records the values of $\varphi$, $t$ and $v$ and marks `id` as open for registration in the next $t_{reg}$ units of time.

**Step 2: Participant Registration.** Whenever there is an open call for participation in generating a random bit with a given `id`, anyone on the network can participate in the process by registering with the RBGC. To register, a participant $p$ should generate a random bit $b_p$, and a nonce $n_p$. She should then compute the hash value $h_p = \text{hash}(b_p, n_p, p, \text{id})$ using a predefined hash function. Then, she can call the `register` function of the RBGC with the following parameters:
- The request identification number `id`.
- A deposit of $v$ units.
- The hash value $h_p$.

This function registers the values of $p$ and $h_p$ in the RBGC. We take this to mean that participant $p$ has committed to providing a bit and a nonce, whose joint hash value with $p$ and the `id` must be equal to $h_p$. The `register` function also numbers the participants from 1 to $n$.

**Step 3: Revealing Choices.** RBGC enters this step as soon as the $t_{reg}$ units of time allocated for Step 2 are passed. This step continues until the deadline $t$, set by the client back in Step 1.

In this step, no new registrations are accepted. Instead, any participant who has already registered can reveal their bit $b_p$ and nonce $n_p$ by calling the `reveal` function of the RBGC with parameters $b_p$ and $n_p$. If the parameters are invalid, i.e. if hash($b_p$, $n_p$, $p$, `id`) is not equal to $h_p$, then the call is ignored. Also, `reveal` keeps track of the following 5 values:

- $n'$: number of participants $p$ who correctly revealed $b_p$
- $n_0$: number of *even*-numbered participants $p$ with $b_p = 0$
- $n_1$: number of *odd*-numbered participants $p$ with $b_p = 0$
- $n_2$: number of *even*-numbered participants $p$ with $b_p = 1$
- $n_3$: number of *odd*-numbered participants $p$ with $b_p = 1$

Intuitively, this construction is similar to the translation between strategies and bits in an RBG game.

**Step 4: Returning Deposits.** After the deadline $t$, each participant can call the `returnDeposit` function of the RBGC. This function checks that the participant has revealed her choice correctly and in time, and returns the participant's deposit only if the check passes.

**Step 5: The RBG Game.** After the deadline $t$, each participant can call the `requestReward` function of the RBGC. This function uses the fee $\varphi$, paid by the client, to reward the participants for submitting inputs. The amount $\varphi$ is distributed among the participants who have correctly revealed their choices in Step 3 as a reward.

We use an RBG game $G$ with $n'$ players to define each participant's share of the reward. In $G$, a participant counted in $n_i$ in Step 3 is assumed to have played the strategy $i$. We denote the resulting outcome of the game by $s$. Let $\alpha = \varphi / n'$, then each participant $p$ receives a reward of $r_p := \alpha \cdot \left(1 + u_p(s)/n'\right)$ by calling the `requestReward` function. Intuitively, a participants' reward is directly dependent on her utility in the RBG game $G$ and the total reward paid to all participants is equal to $\varphi$. Note that the rewards can be computed using the five values tracked in Step 3. In the RBG, a player $p$ who played strategy $i$ has utility $u_p(s) = n_{i-1 (\text{mod} 4)} - n_{i+1 (\text{mod} 4)}$.

**Step 6: Returning the Output.** Finally, the client can request her randomly generated bit by calling the `getOutput` function of the RBGC after the deadline $t$. In this case, the output random bit and return value are decided as follows:
- If no one has participated or no participant has revealed her bit, a *failure* result is returned to the client, together with a refund of the fee $\varphi$.
- Otherwise, the output random bit that is returned to the client is the XOR of all the bits that were correctly revealed by the participants in Step 3.
  - If every participant has revealed her bit correctly, a *success* result will be returned to the client.
  - Otherwise, a *penalty* result will be returned and the confiscated deposits will be paid to the client.

Note that the RBGC can handle multiple random number generation tasks in parallel and these tasks are independent and need not be synchronized.

IX. REQUIREMENTS ANALYSIS

We now prove that our RBGC approach satisfies all the requirements specified in Section VI and is therefore a secure, unmanipulatable, and correctly-incentivized approach for generating random numbers on the blockchain.

---

[3] Generating a multibit random number is no more complicated than generating a single random bit, e.g. one can generate random 32-bit integers by running 32 parallel instances, i.e. asking each participant to submit a 32-bit random number and then running an independent RBG game for each bit.

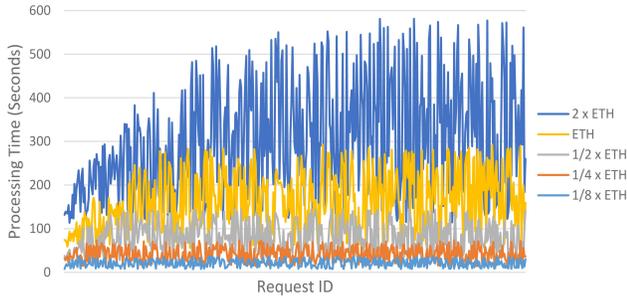

**Figure 3.** Processing times of Requests to RBGC.

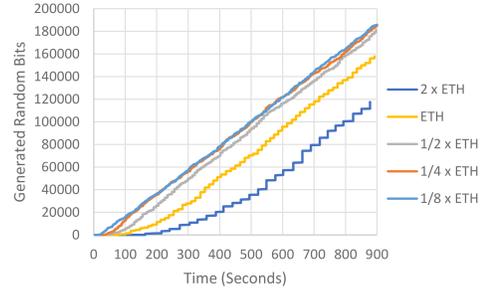

**Figure 4.** Throughput of the RBGC over time.

***Lemma 1 (Functional Requirements.)*** The RBGC approach satisfies the Functional Requirements of Section VI.
*Proof.* (a) All steps of RBGC are implementable as a smart contract in any Turing-complete programmable blockchain. We are also providing a proof-of-concept implementation in Ethereum (Section X). (b) Other contracts can use it as a library by calling the requestRandomBit function, and (c) reliance on participants is achieved by design. ∎

***Lemma 2 (Security Requirements.)*** The RBGC approach satisfies the Security Requirements of Section VI.
*Proof.* Requirement (a) is satisfied by Step 6. Note that when a *penalty* result happens, at least one deposit of *v* units is confiscated and paid to the client, and when a *failure* result happens, the fee *φ* has not been used as a reward and is refunded to the client. (b) The final output bit is the XOR of all the bits revealed by the participants, hence even if one of the participants submits a uniformly randomly generated bit, the overall result is also uniformly random. (c) Openness is achieved by design in Step 2. (d) The final result has no dependence on any value that can be controlled by the miners, hence they cannot tamper with it. Moreover, if the time allocated for Step 3 is long enough, no miner can bar a participant from revealing her choice by dropping a block, because the choice can be revealed in the next block by another miner. If this period is *m* blocks long, then a malicious miner, who aims to stop the revelation from appearing on the blockchain, must mine all the *m* consecutive blocks of this phase, which is very unlikely[4]. (e) RBGC returns *success* only if every submitted bit was correctly revealed, i.e. only if there was no tampering. (f) RBGC generates a new random bit for each request and does not reuse the outputs. ∎

***Lemma 3 (Incentive Requirements.)*** The RBGC approach satisfies the Incentive Requirements of Section VI.
*Proof.* (a) If a participant tries to manipulate the output, i.e. if he does not reveal his bit correctly, his deposit of *v* units will be confiscated (Step 4) and the result of the process changes from *success* to *penalty* (Step 6). (b) Participants' rewards directly correspond to their utilities in an RBG game (Step 5). By Theorem 1, the unique quasi-strong equilibrium of an RBG is when every player chooses a uniformly random bit[5]. ∎

---

[4] A basic assumption in proof-of-work blockchains is that more than half of the computational power is controlled by honest miners. Using this assumption, a revelation can be withheld with a probability of at most $2^{-m}$.

[5] Note that *uniqueness* is not part of the requirements. Our approach surpasses the requirements by providing this stronger guarantee.

***Theorem 2 (Soundness.)*** The RBGC approach of Section VIII is a correct approach for generating secure unmanipulable random numbers on the blockchain. Moreover, it correctly incentivizes the participants to honestly submit random inputs.
*Proof.* This is a direct consequence of Lemmas 1-3 above, which show that the RBGC approach satisfies all the necessary requirements as formalized in Section VI. ∎

### X. Implementation and Experimental Evaluation

***Implementation.*** We implemented the RBGC in Solidity. The code is available at ist.ac.at/~akafshda/rbgc. A noteworthy point about our implementation is that it is entirely loop-free and all of its functions have constant runtime and gas usage.

***Experiment Specifications.*** We simulated the RBGC in a local Ethereum blockchain using Go Ethereum (geth) [70]. In our experiments, 256 random bits were requested every second. We experimented with different block mining difficulties, which led to different rates for the creation of new blocks. Specifically, we experimented with the Ethereum difficulty (ETH), in which the average block generation time $t_{gen}$ is 14133 milliseconds, as well as twice, half, one fourth and one eighth of the default mining difficulty in Ethereum. In each experiment, we set $t_{reg}$ (the time for Step 2) to 3 times $t_{gen}$ and $t_{min}$ (the overall time for generating a random bit) to 10 times $t_{gen}$. The results were obtained on an Intel Core i5-2520M dual-core (2.5 GHz) machine running Microsoft Windows 10.

***Experimental Results.*** Figure 3 shows the processing time of each request to the RBGC, i.e. the time from Step 1 to the end of Step 6, and Figure 4 shows the throughput, i.e. number of successfully generated random bits as time goes by.

***Scalability.*** In general, our approach is very scalable. Our experiments show that on blockchains with the same mining difficulty as Ethereum, RBGC has a throughput of 176.8 random bits per second. Note that the processing times are heavily dependent on block generation times. In contrast, the throughput shows much less dependence, because the RBGC can efficiently handle many requests in parallel.

### XI. Conclusion

In this work, we formalized the functional and security requirements for random number generation on the blockchain and provided the first provably secure, well-incentivized and unmanipulable approach for this problem. We implemented our approach in Solidity and, through experimental results, showed that it is scalable and has a high throughput.

XII. ACKNOWLEDGMENTS

We are very thankful to the anonymous reviewers for their insightful comments that significantly improved the present paper. The research was partially supported by Vienna Science and Technology Fund (WWTF) Project ICT15-003, Austrian Science Fund (FWF) NFN Grant No S11407-N23 (RiSE/SHiNE), ERC Starting Grant (279307: Graph Games), an IBM PhD Fellowship, and a DOC Fellowship of the Austrian Academy of Sciences (ÖAW).